\documentclass[aps,pre,twocolumn,showpacs,groupedaddress]{revtex4-2}

\usepackage{amssymb,amsmath,graphicx,bm,textcomp,xcolor,commath,bm}
\usepackage{ulem}
\usepackage[utf8]{inputenc}
\usepackage{algorithm}
\usepackage{algpseudocode}

\def \bfr {{\bf r}}
\def \bfn {{\bf n}}
\def \bfe {{\bf e}}
\def \bfu {{\bf u}}
\def \bfv {{\bf v}}
\def \bff {{\bf f}}
\def \bfg {{\bf g}}
\def \bft {{\bf t}}
\def \bfk {{\bf k}}
\def \bfl {{\bf l}}

\def \bfa {{\bf a}}
\def \bfm {{\bf m}}
\def \bfT {{\bf T}}
\def \bfN {{\bf N}}
\def \bfB {{\bf B}}
\def \bftau {{\bf \tau}}
\def \bfOmega {{\bf \Omega}}
\def \HT {\mathcal{H}}

\begin{document}

\title{A Discrete Element Method model for frictional fibers}

\author{J\'er\^ome Crassous} \affiliation{Univ Rennes, CNRS, IPR (Institut de Physique de Rennes) - UMR 6251, F-35000 Rennes, France}
\email{jerome.crassous@univ-rennes1.fr}

\date{\today}

\begin{abstract}
We present a Discrete Element Method algorithm for the simulation of elastic fibers in frictional contacts. The fibers are modeled as chains of cylindrical segments connected to each other by springs taking into account elongation, bending and torsion forces. The frictional contacts between the cylinders are modeled using a Cundall and Strack model routinely used in granular material simulations. The physical scales for simulations, the determination and the tracking of contacts, and the algorithm are discussed. Tests on different situations involving few or many contact points are presented and compared to experiments or to theoretical predictions.
\end{abstract}

\maketitle

\section{introduction}

The use of natural or artificial fibers allows to design materials with original mechanical properties. At the nanometric or micrometric scales, carbon nanotubes~\cite{vigolo.2000} or  polymer fibers~\cite{frenot.2003} can be assembled into threads or networks. At the micrometer and millimeter scales, the frictional forces act with the elasticity of the fibers to produce a wide variety of materials. The fibers can just be deposited without any special preparation to form highly elastic media~\cite{toll.1998} such as cushions or non-woven fabrics~\cite{picu.2019}. Textile fibers can be twisted to produce yarns~\cite{pan.2014,warren.2018,seguin.2022}, which are then assembled into cords~\cite{Bohr.2011}, woven~\cite{hearle.1969.book} or knitted fabrics~\cite{hearle.1969.book,poincloux.2018b}. Cyclic mechanical stresses can form very compact natural structures~\cite{verhille.2017}, and birds also assemble fibers to build their nests~\cite{andrade.2021,weiner.2020}. The contacts between fibers play a fundamental role in describing the physics of knots, which is a subtle competition between tension and friction~\cite{bayman.1977,jawed.2015}, as well as eventual bending of the fibers~\cite{audoly.2007,audoly.2009,grandgeorge.2021,johanns.2021}.

Several approaches have been proposed to numerically simulate these structures. One approach is to use finite element algorithms to discretize the fibers~\cite{baek.2020}.  This approach allows a complete solution of the elasticity equations in complex geometries such as nodes~\cite{grandgeorge.2021}, but is only possible for systems with small numbers of contacts. Another approach is to model the fibers as connected spheres~\cite{tangri.2017} or sphero-cylinders~\cite{langston.2015} and to use the discrete element method algorithm widely used for the study of granular materials. However, the periodic variations of diameter of such fiber may induce very specific physical properties as interlocked granular chains stiffening~\cite{dumont.2018}.

More realistic approaches are the simulations of fibers as discrete~\cite{bergou.2008} or continuous~\cite{durville.2010,durville.2012} cylindrical elastic chains of circular cross-sections. The non-interpenetration condition between fibers and surfaces, or between fibers, is then treated as constraints on the displacements. The introduction of frictional tangential forces in such model has been proposed using methods for finding forces that match the Coulomb conditions~\cite{durville.2010,durville.2012,bertails.2011}. In those algorithms, the fibers are moved in order to find the positions of the surfaces that match the non-penetration of fibers, with forces verifying the Coulomb condition. Those positions are found using an iterative procedure with proper regularization of Coulomb law to ensure the convergence towards one solution verifying the force balance. In the case where many frictional contacts are present, the problem becomes hyperstatic, and the solution is expected not to be unique. This is a well known situation in granular material~\cite{moreau.2004} simulations, and the solutions selected by iterative algorithms are not well controlled~\cite{moreau.2004}, and presumably depend on the algorithm itself. Those drawbacks are of course of minimal importance in situations where the indeterminacy in contact forces is absent (hypo- or iso-static problem) such as in knots with few contacts~\cite{choi.2021}, or if qualitative simulations are needed as in computer graphic community~\cite{ly.2020}. In explicit methods, the forces are obtained directly from the kinematic of the body in contacts. The selection of one solution of the Coulomb friction forces among many ones is then ensured by the dynamics of the system. In counterpart, explicit algorithm are usually slower.

Chains of cylinders with frictional contacts have been first introduced in Discrete Element Method by Chareyre~{\it et al.}. These authors used them for the study of the mechanical properties of granular materials reinforced with fibers~\cite{chareyre.2005,bourrier.2013}, with geotextiles~\cite{effeindzourou.2016}, and for the behavior of suspensions of frictional fibers in viscous flow~\cite{kunhappan.2017}.

This bibliography shows that the modeling of fibers in chains of discrete elements has been the subject of many studies, but scattered in different fields. Moreover, the ability of these different models to quantitatively reproduce the behavior of fibers systems with many frictional contacts has never been shown.  Systems of fibers in frictional interactions are the object of a growing interest of physicists and mechanics. The object of this study is to propose to the community a simple Discrete Element Method, easily reproducible, and founded on the Discrete Element Rod model which include frictional contacts, and whose capacity to reproduce the behavior of various frictional fibers is clearly demonstrated.

We will base the model on the theory of elastic chains as proposed by Bergou~{\it et al.}~\cite{bergou.2008}. We will keep a formulation with independent elastic constants of torsion, bending, and torsion, i.e. not linked by a cylindrical beam elasticity. This will allow to simulate various systems, such as arbitrarily flexible wires. The contacts will be treated following an approach proposed by Chareyre~{\it et al.}~\cite{chareyre.2005}. The ingredients of the modeling, as well as the calculations, will be presented in the simplest possible way so that this simulation can easily be reproduced by physicists from various fields.

The manuscript is organized in the following way. In the section II, we first describe the mechanical model of our fibers, including  internal elastic forces and contact forces. The numerical resolution is then detailed in section III, where we insist on points that are specific compared to  DEM simulations of frictional beads, i.e. the numerical scales that are used, the integration of displacement, and the search of neighbors. In section IV, we illustrate this algorithm on various situations including static and dynamics, with few and many contacts.

\section{Mechanical model of fibers in contact}

\subsection{Description of the fiber}

\begin{figure}[t]
\includegraphics[width=\columnwidth]{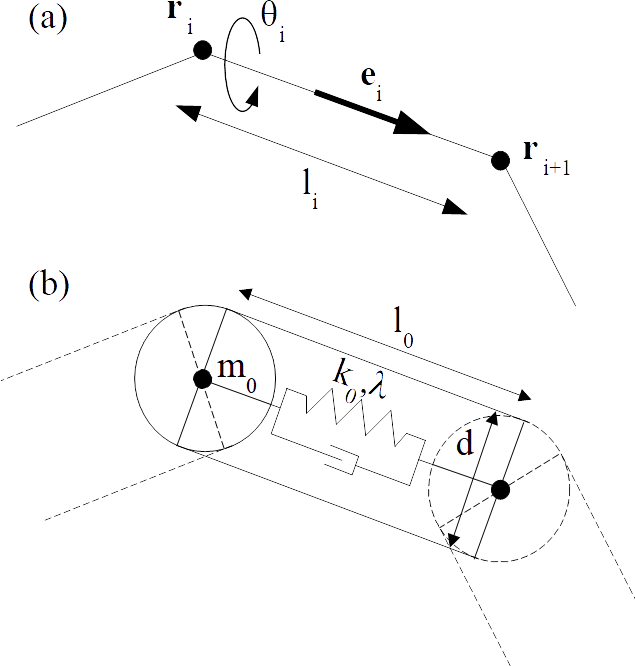}
\caption{(a) Ensemble of connected point forming the skeleton of the fiber. (b) Cylinders and spheres forming the shell of the fibers. }
\label{fig1}
\end{figure}

Following~\cite{bergou.2008}, we model a fiber as an ensemble of $N$ connected points (see figure~\ref{fig1}(a)). Let $\bfr_i$, with $0 \le i \le N-1$ be the position of the point, and $\bfe_i=(\bfr_{i+1}-\bfr_i)/\| \bfr_{i+1}-\bfr_i \|$, with $0 \le i \le N-2$ the unit vector joining two successive points. We note $l_i=\|\bfr_{i+1}-\bfr_i \|$. The segment joining two successive points is the generatrix of a cylinder of circular basis of diameter $d$. In addition, each point $\bfr_i$ is the center of a sphere of diameter $d$. So each fiber is a set of $N$ spheres connected by $N-1$ cylindrical segments. A mass $m_0$ is assigned to each node of the string, and a moment of inertia $J$ is assigned to each cylinder.

The kinematic of the deformation is the following. The different nodes of one fibre may translate, allowing the bending and the stretching of the fibre. The cylinders joining the different nodes stay straight cylinders and are not bent when the fiber is deformed. The rectilinear shape allows to determine the contacts between fibers as contacts between cylinders.The cylinders may rotate around their axis, allowing the twist of the fibers. The kinematic of the chain is then determined by the set of $N$ node positions $\bfr_i$, and $N-1$ cylinder rotations $\theta_i$. To those degrees of freedom, we associate forces that act on nodes, and torques along the axis of cylinders. Any system of forces or torques, such as contact forces or elastic forces, acting on a cylinder will be decomposed as an axial torque and forces on nodes. This decomposition will be detailed below for elastic twist torques and contact forces.

\subsection{Internal elastic forces}\label{sec.internal.elsatic.forces}
The internal elastic forces that we consider in the following are elongation, flexion and twist forces. The elongational forces are modeled using springs of stiffness $k_0$ with dashpots of damping $\lambda$. The equilibrium length of the spring is $l_0$, and the elongation force exerted by point $i+1$ on the mass located at $\bfr_i$ is:
\begin{equation}
\bff_{i+1;i}^{(e)}=\bigl[k_0~(l_i-l_0) + \lambda~\dot{l_i}\bigr]\bfe_i\label{eq.tension}
\end{equation}
Each point $i$ is submitted to forces from points $i-1$ and $i+1$ so that $\bff_{i}^{(e)}=\bff_{i+1;i}^{(e)}+\bff_{i-1;i}^{(e)}$, excepted the first $i=0$ and last $i=N-1$ points.

The flexion forces acting on the point $i$ is obtained from the elastic bending energy $E^{(b)}=\int_s (B/2)~\kappa^2~ds$ with $B$ the bending stiffness of the fiber, and $\kappa$ the curvature. The bending energy of the discrete fiber is:
\begin{equation}
E^{(b)}=\frac{B~l_0}{2}\sum_{i=1}^{i=N-2} \kappa_i^2 \label{eqEb}
\end{equation}
where $\kappa_i$ is the curvature at node $i$, and the summation is extended to all nodes except ending ones. Writing the curvatures $\kappa_i$ as function of nodes positions $\bfr_i$, the flexion force $\bff_{i}^{(b)}=-(\partial E^{(b)}/\partial \bfr_i)$ acting on nodes $i$ is (see Appendix A):
\begin{equation}
\bff_{i}^{(b)}=-\frac{ B}{l_0^3}\bigl[\bfr_{i-2}-4\bfr_{i-1}+6\bfr_{i}-4\bfr_{i+1}+\bfr_{i+2}\bigr]\label{eq.fflexion}
\end{equation}
for $(N-3)\ge i \ge 2$. Expressions of the forces $\bff_{i}^{(b)}$ for $i<2$ and $i>(N-3)$ are given in Appendix A. The calculation supposes that the fibers are weakly extended and bent (see Appendix A).

The internal elastic torque is obtained from the twisting energy~\cite{bergou.2008}:
 $E^{(t)}=\int_s (C/2)~\tau^2~ds$ with $C$ the torsion modulus of the fiber, and $\tau$ the twist of the fiber. The twist may be written as~\cite{love.1920,langer.1996,vanderHeijden.2000}: $\tau=\tau_{int}+\tau_s$. The internal twist $\tau_{int}$ is the twist of the fiber if simply unbent, whereas $\tau_s$ is the torsion of the fibre centre line. Writing internal twist at node $i$ as $(\theta_i-\theta_{i-1})/l_0$, and $\tau_{s,i}$ the torsion of the center line at node $i$, we obtain the twist energy of the discrete fiber as:
\begin{equation}
E^{(t)}=\frac{C~l_0}{2}\sum_{i=1}^{i=N-2}
\bigl(\theta_i-\theta_{i-1}+l_0~\tau_{s,i}\bigr)^2
\label{eqEt}
\end{equation}

The twist moment acting on segment joining nodes $i$ and $i+1$ is obtained by differentiating~\eqref{eqEt} with respect to $\theta_i$~(see \ref{appendix.twist}):
\begin{equation}
 \begin{split}
\bfm^{(t)}_{i}=\frac{C}{l_0}
\bigl[&(\theta_{i+1}-\theta_i+l_0~\tau_{s,i+1})~\bfe_{i+1}\\
-&(\theta_{i}-\theta_{i-1}-l_0~\tau_{s,i})~\bfe_{i-1}\bigr]
\label{eq.Ener.twist}
\end{split}
\end{equation}

This elastic moment is split into one component $m^{(t)}_{i}=\bfm^{(t)}_{i}\cdot\bfe_i$ of the moment along the axis of the segment $(i,i+1)$, and into forces acting on nodes (see \ref{appendix.twist}).

\subsection{Contact forces}

\begin{figure}[t]
\includegraphics[width=\columnwidth]{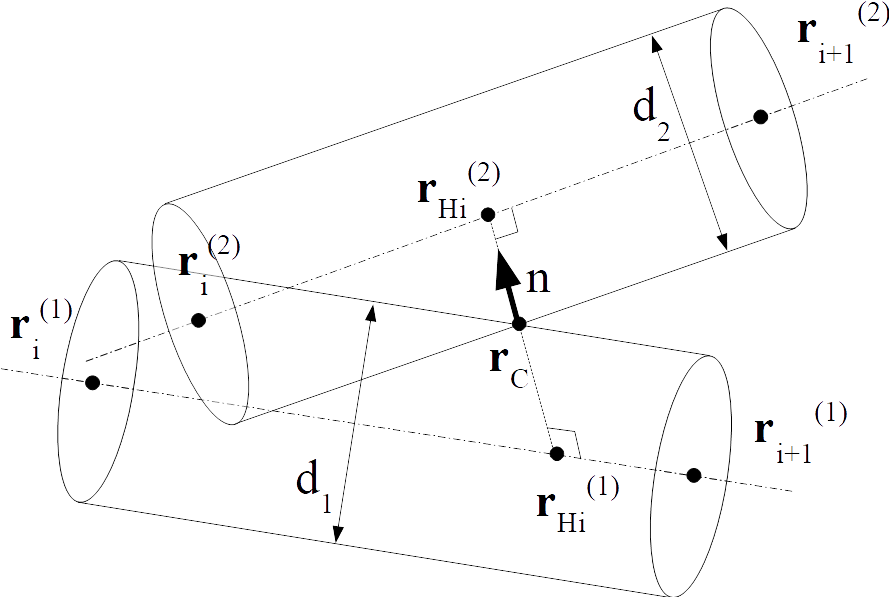}
\caption{Contact between two cylinders.}
\label{fig.contact}
\end{figure}

Contact between fibers may occur between segments of cylinders or spheres belonging to identical or different fibers. The figure~\ref{fig.contact} shows the contact between two sections of cylinders. The contact point $\bfr_C(t)$ is located on the segment with ending points $\bfr_{H_{i}}^{(1)}$ and $\bfr_{H_{i}}^{(2)}$ on axis cylinders which minimizes the distance between axis. This segment is unique if the axis are not parallel. The determination of this segment will be detailed in section~\ref{sec.num.col}. Let $d(t)=\| \bfr_{H_{i}}^{(2)}-\bfr_{H_{i}}^{(1)}\|$ this minimal distance, and $\bfn(t)$ the normal unitary vector. We note $\delta(t)=(d_1+d_2)/2-d(t)$ the interpenetration of the two cylinders, and  $\bfr_C(t)={\bfr_{ H_{i}}^{(1)}}+(r_1-\delta/2)~\bfn$ the contact point. We use the Cundall-Strack model for the contact force~\cite{cundall.1979}. The normal contact force exerted by cylinder 1 on cylinder 2 is modeled as a spring-dashpot system:

\begin{equation}
\bff^{(c)}_n=-\bigl[ k_n~\delta + \lambda_n~\dot{\delta} \bigr] \bfn \label{eq.fn}
\end{equation}

with $k_n$ the contact stiffness between the cylinders, and $\lambda_n$ the contact damping. This contact law is a simplified version of the elastic contact force between 2 cylinders~\cite{timoshenko} which varies non-linearly with the interpenetration $f^{(c)}_n\sim \delta^{3/2}$, and which depends on the angle between cylinder axis. The tangential contact force  is a Coulomb-Force:

\begin{equation}
\bff^{(c)}_t=-Min\bigl[k_t~u_t; \mu k_n \delta\bigr]~\frac{\bfu_t}{u_t} \label{eq.ft}
\end{equation}

where $k_t$ is the tangential stiffness, $\bfu_t$ the tangential displacement, and $\mu$ the microscopic friction coefficient. The tangential displacement $\bfu_t$ is initialized to $0$ when the contact is first formed, and is evolves with time in the following way:

First, since the tangential displacement is expressed in the fixed frame, $\bfu_t$ is first rotated to take into account the rotation of the normal vector. Lets $d\beta$ the angle between the normal at times $t-dt$ and $t$: $\bfn(t-dt)$ and $\bfn(t)$, and $\bfk=\bigl[\bfn(t-dt)\times \bfn(t)\bigr]/\rvert \bfn(t-dt)\times \bfn(t) \lvert$ the axis rotation. We name $\bfu^{rot}_t(t-dt)$ this rotated tangential displacement.

Then, the displacement is integrated as:
\begin{equation}
    \bfu_t(t)=\bfu^{rot}_t(t-dt)+(\bfv^{(2)}-\bfv^{(1)})dt
\end{equation}
where $\bfv^{(1)}$ (and similar for $\bfv^{(2)}$) is the velocity of the point of the cylinder $(1)$ coinciding with the contact point $C$. The velocity is $\bfv^{(1)}=\dot{\bfr}_i^{(1)}+ \bfOmega_i^{(1)} \times (\bfr_C-\bfr_i^{(1)})$ with $\bfOmega_i^{(1)}$ the rotational velocity of the segment $i$ of fiber $1$. The rotation vector is separated into an axial and non-axial components as: $\bfOmega^{(1)}= \bfOmega^{(1)}_\perp + \dot{\theta}_i \bfe_i$, where the non-axial component is $\bfOmega_{i,\perp}^{(1)}=\frac{1}{l_i}\bfe_i\times(\dot{\bfr}_{i+1}^{(1)}-\dot{\bfr}_{i}^{(1)})$.

Finally, the normal component  $(\bfu_t\cdot \bfn)~\bfn$ is removed. If $k_t u_t  > \mu k_n \delta$, then the tangential displacement is renormalized such that $u_t = \mu k_n \delta /k_t$.

The contact force $\bff^{(c)}$ is then expressed as a system of forces $\bff_i^{(c)}$ and $\bff_{i+1}^{(c)}$ applied on nodes $i$ and $i+1$, and a moment $m_i\bfe_i$ acting the cylinder connecting those nodes. The conservation of the resultant and of the moment of contact force implies that:
\begin{subequations}
\begin{align}
\bff_i^{(c)}+\bff_{i+1}^{(c)}&=\bff^{(c)}\\
(\bfr_{i+1}-\bfr_i)\times\bff_{i+1}+m_i\bfe_i&=(\bfr_{C}-\bfr_i)\times\bff^{(c)}
\end{align}
\label{eq.split.fc}
\end{subequations}

A possible choice for forces and moment is (see \ref{appendix.fc}):
\begin{subequations}
\begin{align}
m_i^{(c)}&=\bigl[(\bfr_C-\bfr_i)\times\bff^{(c)})\bigr]\cdot\bfe_i\\
\bff_{i}^{(c)}&=(1-s_i) \bff^{(c)} + \frac{R}{l_i} (\bff^{(c)}\cdot\bfe_i)\bfn\\
\bff_{i+1}^{(c)}&=s_i \bff^{(c)} - \frac{R}{l_i} (\bff^{(c)}\cdot\bfe_i)\bfn
\end{align}
\label{eq.split.fc2}
\end{subequations}
It should be noticed that \eqref{eq.split.fc} does not set all the components of $\bff_{i}^{(c)}$ and $\bff_{i+1}^{(c)}$, and that a supplementary conditions expressed in appendix~\ref{appendix.fc} must be added to obtain \eqref{eq.split.fc2}.

If the contact between two fibers involve one cylindrical segment of the fiber and one sphere, or two spheres, the contact point is calculated accordingly to the type of the surfaces in contact. The translation velocity of the sphere is the velocity of the node. The rotation velocity of the sphere at node $i$ is the rotation velocity of the cylinder joining nodes $i$ with $i+1$.

\subsection{Miscellaneous forces.}

In addition misc extra forces may be added. A global viscous damping force $\bff_i^{(v)}=-\lambda_v~\dot{\bfr_i}$ may be added. It is useful to damp transverse motion of fibers. Indeed, our mechanical model of fiber does not include any dissipation for motion perpendicular to fiber axis if there is no contacts. Volumetric forces such as gravity forces $\bff_i^{(g)}=m_0~\bfg$ with $\bfg$ the gravity field may be also added. Other external forces may applied to fibers such pre-tension at ends of fibers.

\section{Numerical implementation}

\subsection{Integration of equation of motions}

The dynamical equations of motions writes as:
\begin{subequations}
\begin{align}
M_0~\ddot{\bfr_i}&=\bff_i^{(e)}+\bff_i^{(b)}+\bff_i^{(c)}+\bff_i^{(v)}+\bff_i^{(g)}\\
J_z~\ddot{\theta_i}&=m_i^{(t)}+m_i^{(c)}\label{eq.dyn.torque}
\end{align}
\end{subequations}
The second equation described the rotation of cylinder segment around its axis. We did not consider in~\eqref{eq.dyn.torque} any elastic torque due to torsion of the fiber, and the fiber is free to rotate around node $\bfr_i$. The dynamical equations are integrated using a standard second-order Verlet algorithm~\cite{frenkel.book}.

\subsection{Physical parameters for simulations}

\subsubsection{Physical scales}\label{sec.phys_scales}
We first define mass, length and stiffness scale for the simulation. The mass scale $m_0$ is the point mass of nodes, and the length scale $l_0$ is the equilibrium length of each segment, and the stiffness scale $k_0$ is the elongation stiffness of spring. If all the fibers do not have identical physical properties, those scales are chosen from the fibers of smallest radius. For every physical quantities $x$, with a physical scale $x_0$, we note the non-dimensional quantity as $x^*=x/x_0$.

The time scale is $t_0=(m_0/k_0)^{1/2}$. For fibers of diameters $r=d/2$ made of an elastic material (Young modulus $E$, Poisson coefficient $\nu$) of density $\rho$, we have $k_0=E\pi r^2/l_0$, $m_0=\rho \pi r^2 l_0$, and then $t_0=l_0~(\rho/E)^{1/2}$. The time scale $t_0$ is then the time of propagation of compression waves through one segment of the fiber. The force scale $f_0=k_0~l_0=E\pi r^2$ is the force that extend a hypothetical perfectly elastic fiber by $100\%$.

\subsubsection{Elastic forces and  damping }
When submitted to a traction force $f$, the relative expansion of the fibers is $f/f_0=f^*$. It follows that if we want to stay in the limit of small extension, we should keep $f^*\ll 1$. In practice the simulations are done with $f^*\sim 10^{-5}-10^{-3}$. It should be noted that if $f^*$ is too small, the propagation of transverse waves is very slow when no bending forces are present. Indeed, the velocity of transverse wave $v_t$ in a string of linear density  $\rho_l$ under a tension $f$ is $v_t=(f/\rho_l)^{1/2}$. With $\rho_l=m_0/l_0$, we have $\rho_l^*=1$, and the non-dimensional speed of transverse wave is $v_t^*=(f^*/\rho_l^*)^{1/2}=(f^*)^{1/2}$ when no bending stiffness are present.

The non-dimensional bending stiffness is $B^*=B/k_0 l_0^3$. For an elastic fiber as consider in~\ref{sec.phys_scales}, we have $B=E \pi r^4/4$, and then $B^*=(r^*)^2/4$. Similarly, the non-dimensional torsional modulus is $C^*=C/k_0 l_0^3$. For an elastic fiber of radius $r$, we have $C=E\pi r^4/2(1+\nu)$, and then $C^*=(r^*)^2/2(1+\nu)$.

The longitudinal damping $\lambda$ is chosen to avoid compression waves that travel continuously through the fibers, needing very long time to return to equilibrium. We take $\lambda \sim (k_0 m_0)^{1/2}$, and then $\lambda^* \sim 1$ for this.

\subsubsection{Contact force}
The value of the contact stiffness is fixed from a linearization of the Hertzian contact between two elastic cylinders. If two cylinders of radius $r$, with perpendicular axis are in contact, the problem is equivalent to the  the contact between a sphere of radius $r$ and a plane, and the normal force is $f_n=(4/3)~E_{eff}~r^{1/2}~\delta^{3/2}$, with $E_{eff}=E/(1-\nu^2)$, $\nu$ being the Poisson ratio of the material. For doing the linearization, we arbitrary set that the elastic energy of the Hertzian contact $\sim E_{eff}~r^{1/2}~\delta^{5/2}$ is equal elastic energy $k_n\delta^2/2$ of the spring for a normal force $f$ which is of the order or the traction force that we applied on the fibers. Dropping numerical factor of order $1$, we obtain $k_n=E^{2/3}~f^{1/3}~r^{1/3}$. The non-dimensional stiffness may then be obtain as:
\begin{equation}
k_n^*=\frac{(f^*)^{1/3}}{r^*}
\end{equation}
where we again dropped constant term. $f^*$ is the typical non-dimensional force (i.e. the non-dimensional traction applied to the fibers). This value of $k_n^*$ is a reasonable choice for modeling contact, but evidently different values may be set. In practice, since the tension is of order $f^*\sim 10^{-5}-10^{-3}$, and typical radius are $r^*\sim 10^{-1}$, we have $k_n^*\sim 1$. For sake of simplicity, the tangential stiffness is taken as $k_t^*=k_n^*$.

Some damping of the normal force $\lambda_n$ may be introduced. We took $\lambda_n^* \sim 1$ for rapid relaxation of oscillating motion of contact.

\subsubsection{Time scale for simulation}
The time step $dt$ for simulation is chosen such that the dynamic of length relaxation and of contact establishment is correctly described. The length of segment relaxes on a time scale $\sim (m_0/k_0)^{1/2}=t_0$, whereas the time scale for a contact to establish is $\sim (m_0/k_n)^{1/2}=t_0~(k_0/k_n)^{1/2}$. The time step is chosen as $dt=Min\bigl[t_0;t_0~(k_0/k_n)^{1/2} \bigr]/10$, leading to:
\begin{equation}
dt^*=\frac{1}{10}~Min\bigl[1;(k_n^*)^{-1/2}\bigr]
\end{equation}
such that both relaxat
ions occur on at least $10$ time steps. In practice, since $k_n^*\sim 1$, we take $dt^*=0.1$. For a given set of parameters, it is checked that results are unchanged if time steps are divided by a factor $2$.

\subsection{Computation of contact points} \label{sec.num.col}

The Discrete Element Method is mainly used in assemblies of spherical particles. Due to the anisotropic shape of the segments, our algorithm for the determination of the contact points has some particularities compared to sphere-sphere contact that we discuss in this section.

\subsubsection{Distance between fibers}

\begin{figure}[t]
\includegraphics[width=\columnwidth]{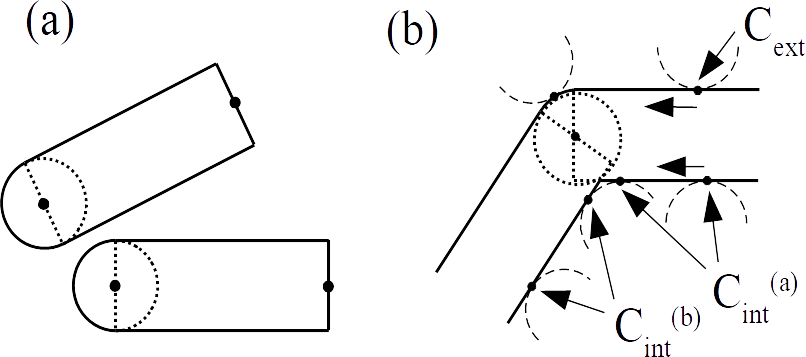}
\caption{(a) 2D view of two sets composed of one sphere and one cylinder. (b) Motions of an external $C_{ext}$ and internal $C_{int}$ contact points at the junction between two cylinders.}
\label{fig.collision}
\end{figure}

The distance between fibers is calculated in the following way. We first consider a segment as a set composed of a sphere and a part of cylinder as shown on figure~\ref{fig.collision}(a).  We first calculate the distance between the two parts of cylinders following the method described in Appendix \ref{appendix.distance}. If contact does not occur along two cylinders, contact between spheres and cylinders are searched, and finally between the two spheres. The hull of the fiber is therefore composed of the external surface of the cylinders and of the spheres as shown ib figure~\ref{fig.collision}(b). The starting and the ending of fibers are finished by spheres.

\subsubsection{Integration of displacement of contact point}
The contact point is followed continuously during the motion of the fibers. This may be done easily as long as the contact point between one segment and one fiber is unique as in example the contact point $C_{ext}$ of ~\ref{fig.collision}(b). In this case, the displacement of the contact point is continuously integrated along the motion. In some case, two contact points may exist simultaneously as the two points as in example the contact points $C_{int}^{(a)}$ and  $C_{int}^{(b)}$ of ~\ref{fig.collision}(b). When the contact at point $C_{int}^{(b)}$ occurs, its tangential displacement is initially set to $0$ (as every new contact), and this lower the tangential force. Since the fibers are weakly bend with $r \ll l_0$, we expect that the number of such contacts are very small compared to the total number of contact, producing very negligible errors. A possible refinement may be to interpolate the two contact points as a single one, allowing a continuous integration of displacement.

\subsubsection{Neighbor search method}

\begin{figure}[t]
\includegraphics[width=\columnwidth]{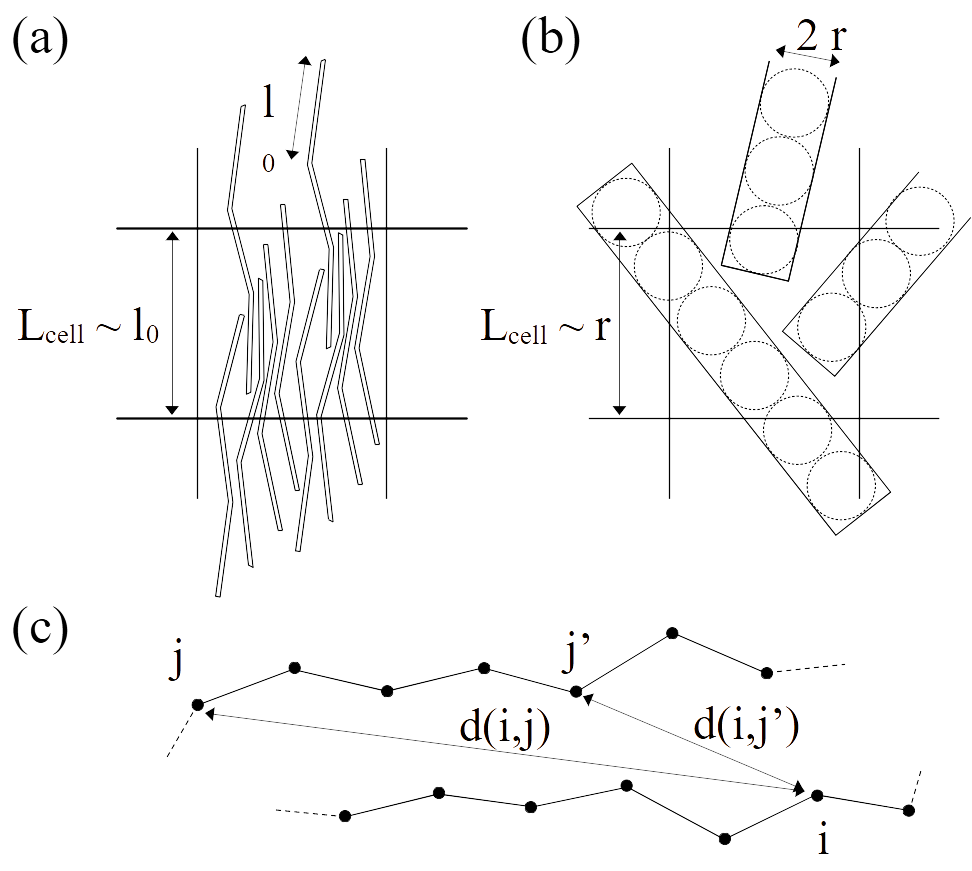}
\caption{(a-b): Two possible choices for the size of the cells into which collisions between fibers mat be searched: (a) size of the cell scales as the length of segments; (b) size of the cell scales as the radius of segments. (c) Simplification arising from the fact that the segments belonging to each fibers are connected.}
\label{fig.cell}
\end{figure}

The search for contacts between discrete objects can significantly increase the computation time of DEM algorithms. In our case, the algorithm for measuring the distance between cylinders is slightly more complex than for spheres, further increasing the computation time of collisions. Several strategies are possible to significantly improve the computation time of the collisions. They are based on the use of neighbor list (Verlet list) or on the partition of the system in boxes (Linked Cell Method). We discuss here the problem arising when using strongly anisotropic objects. In linked cell method, the particles are assigned in cells, and the list of particles is each cell is updated periodically. The collisions are searched only for particles within the same or the neighboring cells. This strategy is very effective for approximately monodisperse spheres. In the case of polydisperse spheres, the size of the cell must be a multiple of the size of the largest particles, so that the number of particles per box increases. As a consequence, the computation time grows rapidly with the polydispersity as shown by Luding {\it et al}~\cite{luding.2007}. The problem is very similar for strongly anistropic particles such fibers, or segments of fibers. The figure~\ref{fig.cell}(a) shows an assembly of fibers with segments of size $l_0$. If collisions between segments are searched within one or neighboring cells, the size of the cell should be $\sim  2~l_0$. For segments of section $\sim 4~r^2$, the number of segments in each cell is $\sim 2~(l_0/r)^2$ for dense $3D$ system. Since $l_0/r \gg 1$, sorting particles in cell of size $\sim l_0$ is not efficient. A more convenient way to define cell may be considered. It consists, as  shown ~\ref{fig.cell}(b), of replacing segments by fictitious spheres of radius $r$ inside each segment of length $l_0$, and to consider cells of size $\sim  4~r$. In this case for a system of $N_f$ fibers of $N$ segments each, the total numbers of fictitious spheres is $\sim N_f N (l_0/2r)$. However those two methods do not use the fact that different segments of one fiber are linked together. Taking advantage of this knowledge may significantly speed up the search of neighboring. The figure~\ref{fig.cell}(c) shows two fibers, and we search contact between segment $i$ of fiber $1$, with fiber $2$, by increasing $j$. For a segment $j$, we calculate the distance $d(i,j)$. If this distance is larger that $2r$ there is no contact, and we are sure that there is no contact between the two fibers for $\vert j'-j \vert \le d(i,j)-2r$. So the next segment where we need to search contact verifies $j' > j+d(i,j)-2r$.

The optimal strategy to find contacts is expected dependent on the type of fiber under studies. In case of fibers with numerous segments, taking advantage of the constraint that the segment are linked as depicted in~\ref{fig.cell}(c) is presumably the better. At the opposite, in the case of an assembly of very short fibers, such as a packing of one-segment needles, use of cell as~\ref{fig.cell}(b) should be preferred. The further study of such optimization is outside the scope of this study.

\section{illustration experiments.}

The program has been tested on various simple geometries in order to check the consistency with the theory, to verify the numerical stability of the algorithm, and test the numerical precision. Those configurations were the  rolling or sliding of a cylinder on a inclined plane, the velocity of transverse waves of a string, the static flexion of a fiber loaded at extremity by a point force, the catenary shape of a massive string under gravity. We present in the following four more complex situations. If not otherwise specified, the simulation parameters are: time step $dt^*=0.1$, internal damping $\lambda^*=2.8$, contact stiffness $k_n^*=k_t^*=1$, contact damping $\lambda_n^*=1$, global viscous damping $\lambda_v^*=0.001$, inertia momentum $J^*=m^*{r^*}^2/2$ (homogeneous cylinder).

\subsection{Elastic rods without contacts}

The elastic rod model has been already tested in misc situations that do not involve frictional contacts~\cite{bergou.2008}. The test examples presented here are just for checking the approximations used in \ref{sec.internal.elsatic.forces} and \ref{sec.appendices}.

\begin{figure}[t]
\includegraphics[width=\columnwidth]{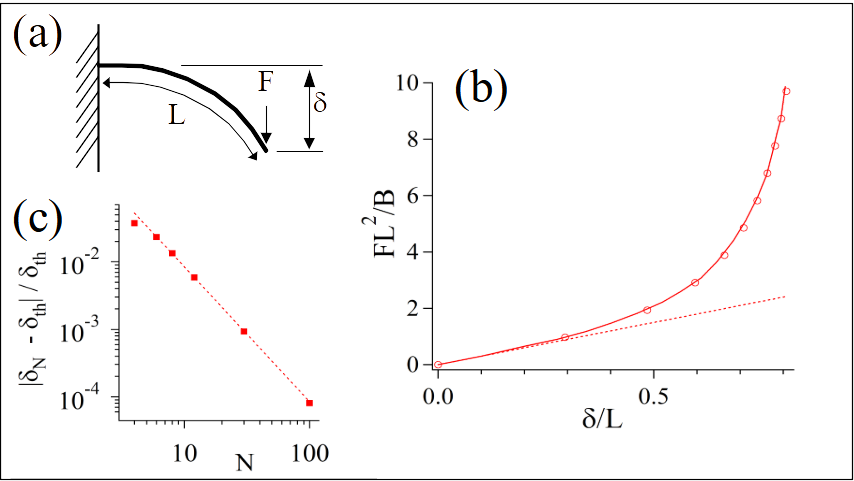}
\caption{Force applied to a beam as a function of its deflection. (a) Geometry. (b) Symbols: simulation results. Plain line: theoretical solution. Dashed line: force in the small deflection limit $FL^2/B=3\delta/L$. (c) Convergence: $\delta_N$ is the end deflection for a beam with $N$ nodes ($(N-1)$ cylinders), and $\delta_{th}$ the theoretical deflection. Applied forces is $FL^2/B=10$. Symbols are relative error, and dashed line is a $N^{-2}$ decay.}
\label{fig.flexion}
\end{figure}

The first example is the deformation of a clamped elastic rod ($N=100$, $B^*=0.1$) submitted to a point force applied at one end (see figure~\ref{fig.flexion}.a). The clamping is imposed by fixing the first and second node of the rod. The free rod length $L$ is then the number of cylinders $N-1$ minus one: $L=N-2=98$. Results for different values of applied forces are shown on~\ref{fig.flexion}.b. Those results may be compared to the deflection of an non-extensible rod. At small deflections $\delta \ll L$,  $FL^2/B \simeq 3\delta/L$. At large deflections $\delta \sim L$ simulations agree well with the analytical solution of Bisshopp et al.~\cite{bisshopp.1945}. Simulations with beams made with different $N$ show that the solution obtained with the discrete beam converges towards analytical solution as $\sim N^{-2}$ (see~fig.\ref{fig.flexion}.c). It may be noticed that since the maximum force applied in those simulations are of order $F\sim 10~B/L^2$, the maximum non-dimensional force $F^*\sim 10^{-4}<<1$, so that the beam stretching is negligible.

\begin{figure}[t]
\includegraphics[width=\columnwidth]{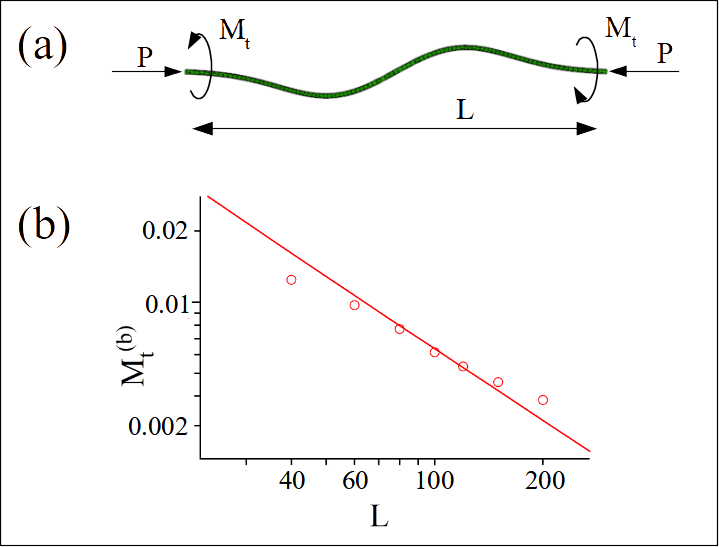}
\caption{Buckling of an elastic rod. (a) Geometry. (b) Torque applied at ends at the torsional buckling threshold as a function of the rod length. Symbol: simulation results with $P=0$. Line: $M_t^{(b)}=2\pi B/L$.}
\label{fig.buckling}
\end{figure}

The second example is the buckling of a rod submitted to a compression and applied torque at its ends (see fig.\ref{fig.buckling}.a). A numeric rod ($B^*=0.1$) is submitted to a torque $M_t$ at its ends. The displacement of the ends perpendicularly to the axis of the rod are blocked, and no compression forces $P$ are applied. The torque is slowly increased until buckling of the beam occurs. The buckling threshold is determined by measuring the displacement of the ends along the axis of the rod. Those displacements are initially negligible, and suddenly increases as buckling occurs. fig.\ref{fig.buckling}.b shows the buckling torque $M_t^{(b)}$ as a function of the rod length. Stability analysis of twisted rods leads to\cite{timoshenko.stability}: $M_t^{(b)}=2\pi B/L$. As shown on fig.\ref{fig.buckling}.b, the numerical results are in correct agreement with this theoretical law.

\subsection{Static without flexion : capstan}
\begin{figure}[t]
\includegraphics[width=\columnwidth]{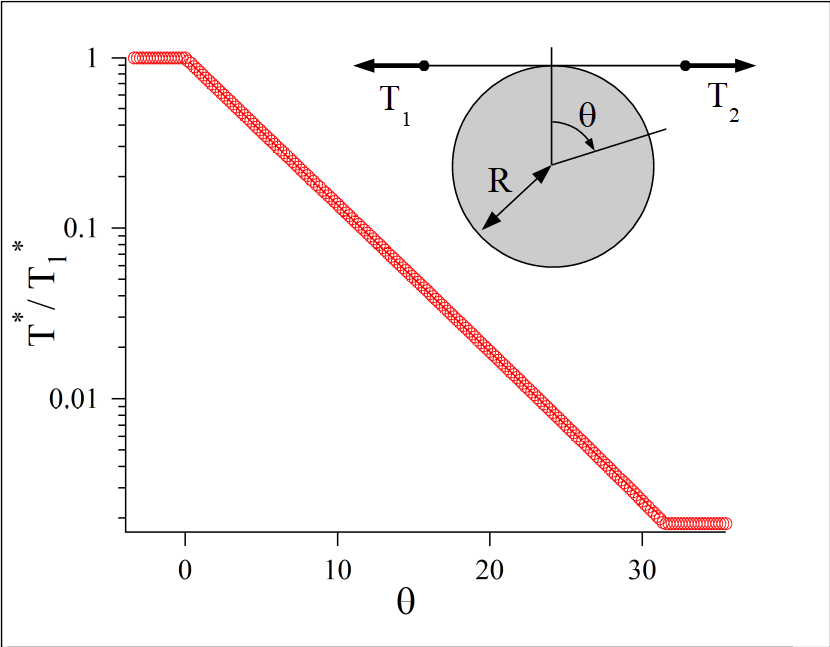}
\caption{Tension in a rolled string around a cylinder: $T^*$ is the tension in the string, and $\theta$ is the rolling angle.Circles are symbol, plain line is an exponential fit. See for simulations parameters Inset: Schematic of the experiment.}
\label{fig.capstan}
\end{figure}
We simulate the tension along a string which is rolled up around a cylinder. For this, we prepare a infinitely flexible spring ($B^*=0$, $N=200$, $r^*=0.1$) which makes $5$ turns around a cylinder ($R^*=5$). The cylinder had a huge mass and moment of inertia to prevent any motion. The friction coefficient is $\mu=0.2$. We first apply an equal tension $T_1^*=T_2^*=0.01$, with opposite directions, to the two ends of the string. We let the system to reach equilibrium. Then, we slowly decreases $T_2^*$ while keeping $T_1^*=0.01$. For a threshold value of $T_2^*$, the  sliding of the string occurs. We measure the tension in the string using~\eqref{eq.tension} at the onset of sliding. The fig.\ref{fig.capstan} shows the decrease of the tension $T^*$ along the string as a function of $\theta=(s^*-s_0^*)/(R^*+r^*)$, with $s^*$ the abscissa along the curve, and $s_0^*$ the abscissa of first contact contact point. The solution of capstan problem with a finite thickness rod predicts that~\cite{jung.2008}: $T^*/T_1^*=\exp(-\mu~\theta)$, which is the observed behavior on figure~\ref{fig.capstan}. The measured decay is $\mu=0.198$ in agreement with the imposed value $\mu=0.2$.

\subsection{Static with flexion : elastic knots}

\begin{figure}[t]
\includegraphics[width=\columnwidth]{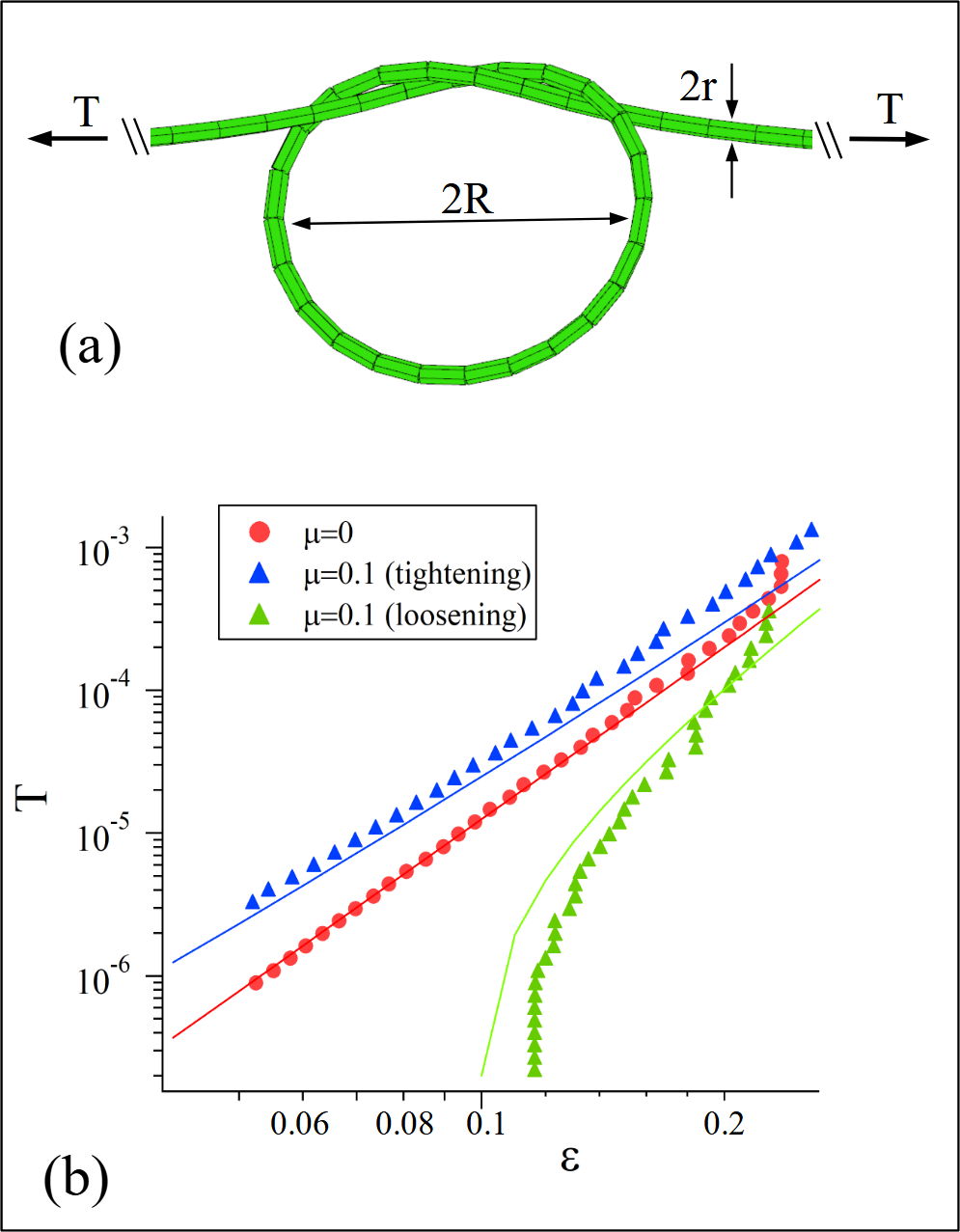}
\caption{(a) Snapshot of a ($3_1$) knot. For seek of clarity, illustration is made with $r^*=0.2$. (b) Tension as a function of $\varepsilon=\sqrt{r/R}$ for frictionless and frictional strings. Symbols are numerical data, and lines are theoretical expressions given by equation~\eqref{eq.audoly}.}
\label{fig.knot}
\end{figure}

We consider the mechanical response of an elastic rod with an open knot. An elastic fiber of length $L$, with a circular section of radius $r$, and bending modulus $B$ is bent in an open trefoil knot ($3_1$). We then apply a tension $T$ to the ends of the fibers. This experimental situation has been addressed by Audoly {\it et al.}~\cite{audoly.2007,audoly.2009}. When the tension is weak, the loop radius $R$ is very large compared to $r$. In this limit, authors found analytical solutions for the shape of the knot, either in the frictionless case, but also for weak friction $\mu \ll 1$. This knot has been simulated very recently by Choi {\it et al.} using an discrete rod model with an implicit solver for the contact force~\cite{choi.2021}.

We simulate numerically such knot by considering a flexible spring ($r^*=0.1$, $B^*=(r^*)^2/4=2.5~10^{-3}$, $N=500$, $\lambda_v^*=4.10^{-4}$) as shown on figure~\ref{fig.knot}(a). We first knot the fiber by setting $\mu=0$ and applying a  tension $\pm~T^*\bfe_z$ at ends. After this preparation stage, we set $\mu$ to its actual value, and we increase or decrease $T^*$ depending if we tighten or lossen the knot. When the knot begins to move, we measure the radius of curvature of the loop as $R=<\|d\bft_i / ds\|^{-1}>$, where $d\bft_i/ds=\bfe_{i+1}-\bfe_{i}$ is the derivative of the tangent vector, and the average $<~>$ is over all segments in the loop which are at a distance of at least one segment from any contact point. Following ~\cite{audoly.2007,audoly.2009}, we introduce $\varepsilon=\sqrt{r/R}$. The figure~\ref{fig.knot}(b) shows the tension $T^*$ as a function of $\varepsilon$ for frictionless ($\mu=0$), and frictional ($\mu=0.1$) loosening and opening knots. The analytical solutions in the limits $\varepsilon\ll1$ and $\mu \ll 1$ are~\cite{audoly.2007,audoly.2009}:
\begin{equation}
\frac{Tr^2}{B}=\frac{\varepsilon^4}{2}\pm\mu\sigma\varepsilon^3
\label{eq.audoly}
\end{equation}
where the sign $\pm$ depends if the knot is tightened $(+)$ or loosened $(-)$, and $\sigma$ is a numerical constant which is $\sigma \simeq 0.492$ for trefoil knot. As shown on figure~\ref{fig.knot}(b), the numerical data agrees correctly with the analytical one. In the frictionless case, we may observe deviations from the scaling $T\sim \varepsilon^4$ when $\varepsilon \gtrsim 0.15$. Two possible sources of deviations may be identified. First, the equation~\eqref{eq.audoly} is obtained in the limit $\varepsilon \ll 1$, and deviations may arise from high order $\varepsilon$ terms in equation~\eqref{eq.audoly}. Second, for $\varepsilon \gtrsim 0.15$, we have $R^*=R/l_0=r^*/\varepsilon ^2\sim 4$, so that the discretization of loop may then be an issue. The discret nature of the rod may be clearly identified on numerical data form $\mu=0.1$ loosening, where some steps in $\varepsilon$ are visible. For the frictional case, the model \eqref{eq.audoly} slightly underestimates the role of friction compared to numerical simulations. It may be due to some departure from the hypothesis $\mu \ll 1$ which is used to obtain \eqref{eq.audoly}.

\subsection{Impact: falling chain.}

\begin{figure}[t]
\includegraphics[width=\columnwidth]{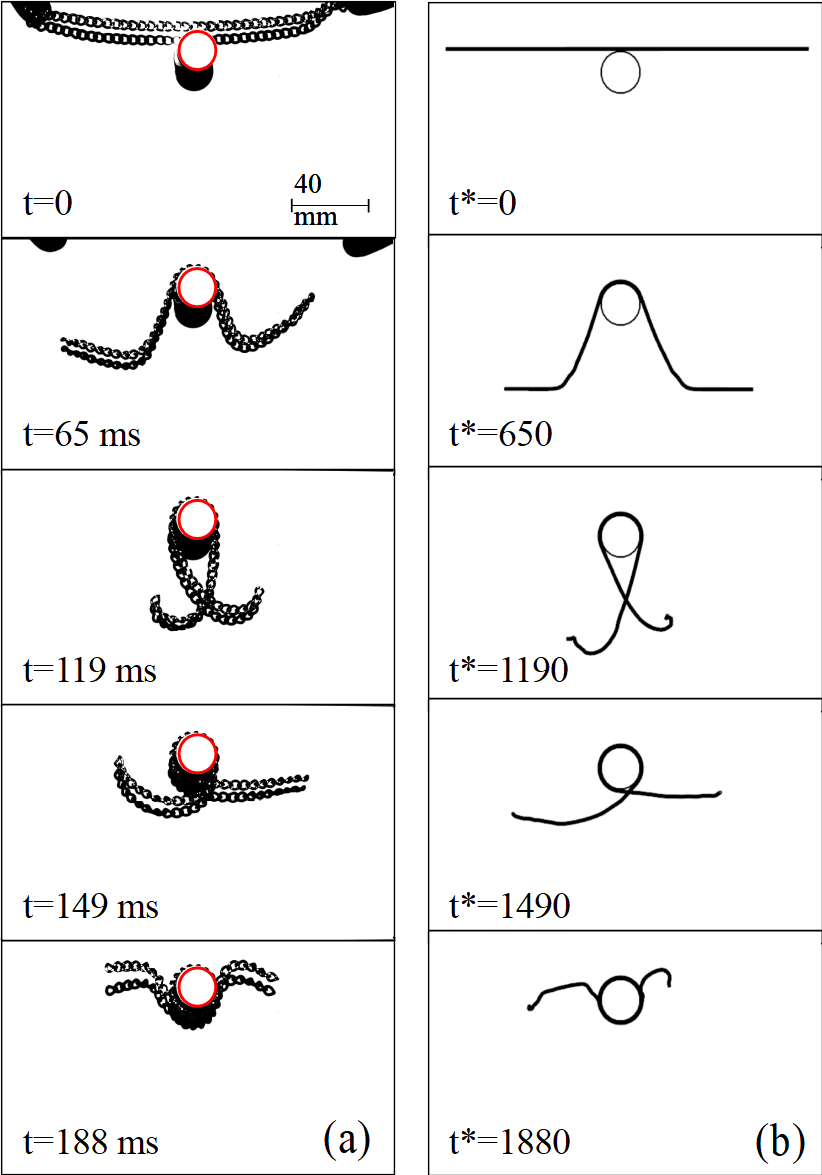}
\caption{(a) Experimental snapshots of the impact of a metallic chain on a fixed perpendicular cylinder of radius $R=10~mm$. The perimeter of the cylinder is underlined in red. Time $t=0$ is defined as the first contact time. Chain length $L=190~mm$, mass $m=8.5~g$. (b) Simulated impact. See text for physical parameters of the simulation.}
\label{fig.fall}
\end{figure}

We consider the dynamics of the impact of a metallic chain on a cylindrical obstacle. We restrict this analysis to a qualitative analysis. A metallic chain (length $L=190~mm$, mass $m=8.5~g$) is held at its extremities by hands. The chain is released and its fall is recorded with a fast camera operating at $200~fps$. The figure \ref{fig.fall}(a) show some snapshots of the impact. The chain is simulated as a infinitely flexible spring $B^*=0$. We set the length scale to $l_0=2~mm$, and $N=95$, so that $L=N~l_0$. The choice of the time scales may be done in the following way. We want to simulate a non-extensible chain, so we need that the non-dimensional typical force is $<<1$.  The gravity force is $F_g=Nm_0g$, with $m_0$ the mass scale of one segment, and $g$ the gravity. The non-dimensional gravity force is then $F_g^*=F_g/k_0 l_0=Ng/l_0t_0^{-2}=Ng^*$. We take $g^*=4.9~10^{-5}$ so that $N~g^*\simeq 5.~10^{-3}\ll 1$. This sets the time scale $t_0=0.1~ms$. It should be noted that in the limit of a non-extensible chain, the mass scale does not need to be specified. Other parameters are $dt^*=0.1$, $R^*=5$, $r^*=0
.1$, $\mu=0.1$, $k_n^*=k_t^*=1$. Figure~\ref{fig.fall}(b) shows the results of the simulations which qualitatively agree with the experiments. We may remark that the behaviour of the experimental chain is not symmetric in compression and in extension (nearly infinite stiffness in extension, zero stiffness in compression), whereas the numerical chain is symmetric (same stiffness in compression and in extension). However, in impact experiment, the chain is always in tension, and the lack of symmetry does not have importance.

\subsection{Multiple fibers: a yarn model.}

\begin{figure}[t]
\includegraphics[width=\columnwidth]{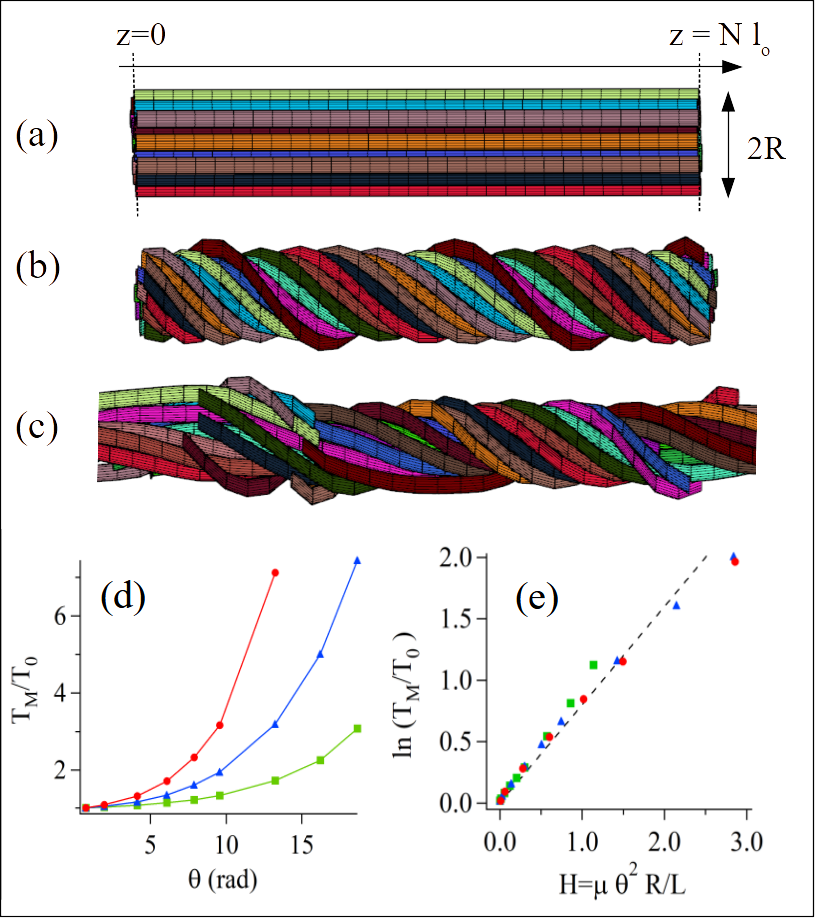}
\caption{(a) Assembly of of initially straight fibers. (b) Thread of fiber after a torque is applied at ends. (c) Separation of the fibers due to applied forces. (d) Force ratio necessary to separate the slivers as a function of the twist angles. Red disks: $\mu=1$, blue triangles: $\mu=0.5$, green squares: $\mu=0.2$. Lines are for guidelines. (e) Same data as (d) plotted as a function of $\HT=\mu \theta^2 R/L$. Dotted line is $0.75~\mu \theta^2 R/L$. Simulation parameters are $N_f=20$, $N=30$, $r^*=0.1$, $B^*=0$. For clarity, the sub-figures (a-c) are enlarged by a factor 6 perpendicularly to $z$-axis.}
\label{fig.yarn}
\end{figure}

In a recent study, Seguin {\it et al.} considered the situation of a staple yarn made of twisted totally flexible fibers~\cite{seguin.2022}. We present in this section some numerical details about this simulation. The yarn is made of an assembly of $N_f$ identical fibers of $N$ segments initially parallel to an axis $z$ (see figure~\ref{fig.yarn}(a). Their positions $(x_i;y_i)$ in the plane perpendicular to the $z$ axis, with $1\le i \le N_f$ are the positions of a packing of disks in 2D obtained from a separate simulation.

In a first phase of the simulation the fibers are twisted. The fiber are submitted to a tension $T^*=10^{-4}$ along $z$, applied at both ends. A torque $C^*\bfe_z$ is applied to both ends of the assembly of fibers. For this, each fiber $i$ with $1\le i \le N_f$ is submitted at both ends $j=0$ and $j=N$ to an external shear force:
\begin{equation}
\bftau_i(j)=\pm\frac{C^*}{\sum_i r_i^2(j)}\bigl[\bfe_z \times \bfr_i(j) \bigr]
\label{eq.torque}
\end{equation}
where sign is $-$ for $j=0$, and $+$ for $j=N$ ends. The torque is gradually increased until it reaches its target value and the shear forces are updated at each time step. Under the action of this torque, the fibers twist and becomes approximately helicoidal as shown on figure~\ref{fig.yarn}(b). During this preparation, the friction coefficient is set to a low value $\mu=0.05$. This is important in order to obtain a regular pitch along the thread. Indeed, since the yarn is twisted by the application of torques at ends, the presence of an important friction between fibers has the effect of concentrating the twist near ends, with a central zone of low twist. This behavior is also observed experimentally~\cite{seguin.2022} if the twist is not homogenized along the yarn. The duration of this preparation stage is $t^*=5.10^{5}$, and the total twist $\theta$ is measured at the end of this phase.

In a second phase the fibers are separated. The friction is first set at its target value. The fibers are randomly partitioned is two set -up- and -down-. The tension of the up-fibers are multiplied by a factor $f>1$ at the up-extremities: $T_{up}(j=0)=T^*$ and $T_{up}(j=N)=f~T^*$.  Symmetrically, $T_{down}(j=0)=f~T^*$ and $T_{down}(j=N)=T^*$. The factor is $f=1$ at the beginning of the separating stage and is increased at a fixed rate $(\Delta f/\Delta t^*)=2.10^{-6}$. During this phase, the torque is kept constant. The difference between the average positions of the -up and -down fibers is measured. This difference stays constant, until a threshold value of $f$ where the two slivers of fiber separates (see figure~\ref{fig.yarn}(c)). A mechanical model of this problem developed in~\cite{seguin.2022}, show that the force necessary to separate the two slivers is $\ln(1+f) \simeq 0.75~\mu \theta^2 R/L$ which is the behavior that is observed on figure~~\ref{fig.yarn}(e).

\section{Conclusion}

We have described a discrete element mechanics algorithm for the simulation of flexible and frictional fibers. This algorithm is similar to the DEM type algorithms widely used for the study of granular materials. The difference arises from the type of surfaces in contact (cylinders and not spheres) and from the elastic forces between the cylinders which are connected to form a fiber. The algorithm has been tested on various configurations that can be compared to  experiments or to theoretical models.

The assumptions and approximations used to design this algorithm are quite limited. The low bending assumption is not very compelling for many applications, but could eventually be minimized by a finer discretization of the fiber. The simplification of the Hertzian elastic contact law between the cylindrical segments by a linear spring has probably a very small impact on the modeling of real systems. An extension to non-linear contact laws should not be a problem. Finally, the discretization of the fiber generates a discontinuity of the displacement for some contact points at the passage between successive segments of a fiber. A priori the number of such jumps is negligible compared to the total number of contacts for thin and weakly bent fibers, and this should not be an issue for simulations of real systems.

The main difference between this algorithm and those previously described to simulate elastic fibers lies in the level of simplification of the mechanical problem. Simulations of fibers with finite element algorithms are certainly of high accuracy but can only simulate small systems. Implicit algorithms are probably faster, but the indermination of forces in multi-contact cases is not resolved by the dynamics of the system. The use of a DEM algorithm is a compromise that allows to consider relatively complex assemblies of fibers and that correctly handles the multiplicity of equilibrium solutions.

The potential applications of this algorithm are obviously multiple. The study of complex knots between fibers of ropes, with or without bending energy is possible. The mechanical response of fiber clusters in nests, cushions, or in rigid needle stacks are also possible. For these studies, the contact search should be optimized according to the aspect ratio of the fibers and the geometry of the packing. The simulation of knitted or woven fabrics can also be considered. For this, large systems can be simulated, but the introduction of periodic boundary conditions should be more suitable. Finally, systems mixing fibers and grains for the study of soils reinforced by fibers or roots are also possible applications of this work.

\begin{acknowledgements}
The author would like to thank Antoine Seguin and Sean McNamara for discussions and careful reading of the manuscript, and Laurent Courbin for his help in experiments on falling chains.
\end{acknowledgements}

\section{Appendices}\label{sec.appendices}

\subsection{Flexion forces}
The curvature $\kappa_i$ at a node $N-2\ge i\ge 1$ is first expressed as a function of the positions of nodes $\bfr_{i-1}$, $\bfr_{i}$ and $\bfr_{i+1}$. The radius $R_i=1/\kappa_i$ of the circle joining those three points may be expressed as a function of the surface $S_i$ and the perimeter $p_i$ of the triangle with vertices $(\bfr_{i-1},\bfr_{i},\bfr_{i+1})$ using the Heron formula. After elementary calculus, we obtain:

\begin{equation}
\kappa_i^2=4~\frac{\bfl_{i-1}^2 \bfl_{i}^2 - (\bfl_{i-1}\cdot \bfl_{i})^2}
{\bfl_{i-1}^2~\bfl_{i}^2~(\bfl_{i-1}+\bfl_{i})^2}\label{eq.A.kappa}
\end{equation}

where we noted $\bfl_i=\bfr_i-\bfr_{i-1}$. The bending energy is
\begin{equation}
E^{(b)}=\frac{B~l_0}{2}\sum_{i=1}^{i=N-2} \kappa_i^2
\end{equation}

The flexion force is then:
\begin{equation}
\bff_{i}^{(b)}=\frac{B~l_0}{2} \frac{\partial} {\partial \bfr_i} \bigl[\sum_{i'=1}^{i'=N-2} \kappa_{i'}^2 \bigr]\label{eq.A.force}
\end{equation}

First, we notice that for weakly bend fibers $\bfl_i \simeq \bfl_{i-1}$, and for weakly extended fibers $l_i \simeq l_0$. Then, the denominator of \eqref{eq.A.kappa} is $\simeq 4~l_0^6$:
\begin{equation}
\frac{\partial \kappa^2_{i'}} {\partial \bfr_i}\simeq
\frac{1} {l_0^6}
\frac{\partial}{\partial \bfr_i}
\bigl[\bfl_{i'-1}^2 \bfl_{i'}^2 - (\bfl_{i'-1}\cdot \bfl_{i'})^2\bigr]
\end{equation}
Using $\bfl_i=\bfr_i-\bfr_{i-1}$, we obtain:
\begin{subequations}
\begin{align}
\frac{\partial \kappa^2_{i-1}} {\partial \bfr_i}&\simeq
\frac{1} {l_0^4} \bigl[2 (\bfl_{i-1}-\bfl_{i-2})\bigr]\\
\frac{\partial \kappa^2_{i}} {\partial \bfr_i}&\simeq
\frac{1} {l_0^4} \bigl[-4 (\bfl_i-\bfl_{i-1})\bigr]\\
\frac{\partial \kappa^2_{i+1}} {\partial \bfr_i}&\simeq
\frac{1} {l_0^4} \bigl[2(\bfl_{i+1}-\bfl_{i})\bigr]
\end{align}
\end{subequations}
and $(\partial \kappa^2_{j}/\partial \bfr_i)=0$ if $\vert i-j \vert >1$. We obtain finally:
\begin{eqnarray}
\bff_{i}^{(b)}&=&\frac{B}{l_0^3} \bigl[-\bfl_{i-2}+3\bfl_{i-1}-3\bfl_{i}+\bfl_{i+1} \bigr]\\
&=&\frac{B}{l_0^3}
\bigl[\bfr_{i-2}-4\bfr_{i-1}+6\bfr_{i}-4\bfr_{i+1}+\bfr_{i+2}\bigr]
\end{eqnarray}
for $N-3\le i \le 2$. Expressions of the forces for $i<2$, and $i>N-3$ are obtained by noticing that summation in \eqref{eq.A.force} is for $i'=1$ to $i'=N-2$.

\begin{subequations}
\begin{align}
\bff_{0}^{(b)}&=-\frac{ B}{l_0^3}\bigl[\bfr_{0}-2\bfr_{1}+\bfr_{2}\bigr]\\
\bff_{1}^{(b)}&=-\frac{ B}{l_0^3}\bigl[-2\bfr_{0}+5\bfr_{1}-4\bfr_{2}+\bfr_{3}\bigr]\\
\bff_{N-2}^{(b)}&=-\frac{ B}{l_0^3}\bigl[\bfr_{N-4}-4\bfr_{N-3}+5\bfr_{N-2}-2\bfr_{N-1}\bigr]\\
\bff_{N-1}^{(b)}&=-\frac{ B}{l_0^3}\bigl[\bfr_{N-3}-2\bfr_{N-2}+\bfr_{N-1}\bigr]
\end{align}
\end{subequations}

\subsection{Twist moment and forces}\label{appendix.twist}

The twist energy of the discrete rod is:
\begin{equation}
E^{(t)}=\frac{C}{2~l_0}\sum_{i=1}^{i=N-2} (\theta_{i}-\theta_{i-1}+l_0~\tau_{s,i})^2 \label{eq.Ener.twist}
\end{equation}
where $\tau_{s,i}$ is the torsion of the center line at node $i$, and $(\theta_{i}-\theta_{i-1})/l_0$ is the internal twist. The torsion $\tau_s$ of the center line is obtained from Frenet-Serret equations as $\tau_s=(d\bfN/ds)\cdot \bfB$, where $(\bfT,\bfN,\bfB)$ are tangent, normal and bi-normal vector of the centre line of the fiber. They are obtained by multiple differentiation of tangent vector $\bfe_i$, with appropriate interpolations depending if derivatives are evaluated at nodes or at cylinder.
\begin{equation}
\bfm^{(t)}_{i-1;i}=\frac{C}{l_0}(\theta_{i}-\theta_{i-1}+l_0~\tau_{s,i})\bfe_{i-1}
\end{equation}
Taking into account the torque acting from the segment $(i+1;i+2)$ on the segment $(i;i+1)$, the total elastic twist torque acting on the segment $(i;i+1)$ is:
\begin{equation}
 \begin{split}
\bfm^{(t)}_{i}=\frac{C}{l_0}
\bigl[&(\theta_{i+1}-\theta_i+l_0~\tau_{s,i+1})~\bfe_{i+1}\\
-&(\theta_{i}-\theta_{i-1}-l_0~\tau_{s,i})~\bfe_{i-1}\bigr]
\label{eq.Ener.twist}
\end{split}
\end{equation}

This torque is split in two components. The axial (co-linear to $\bfe_i$) component is:
\begin{equation}
m^{(t)}_{i}=\bfm^{(t)}_{i} \cdot \bfe_{i}
\end{equation}
whereas the remaining perpendicular component $\bfm^{(t)}_{i} -m^{(t)}_{i} \bfe_{i}$ is written as a system of two points forces $\bff^{(t)}_i$ and
$\bff^{(t)}_{i+1}$ acting at points $i$ and $i+1$ such that:
\begin{subequations}
\begin{align}
\bff^{(t)}_{i}+\bff^{(t)}_{i+1}&=0\label{twist.split.a}\\
(\bfr_{i+1}-\bfr_i) \times \bff^{(t)}_{i+1}&=\bfm^{(t)}_{i} -m^{(t)}_{i} \bfe_{i}\label{twist.split.b}\\
\bff^{(t)}_{i}\cdot \bfe_i&=0\label{twist.split.c}
\end{align}
\end{subequations}
\eqref{twist.split.a} ensures that the system of two points forces is a torque, \eqref{twist.split.b} assigns the moment, and \eqref{twist.split.c} that those forces do not stretch the rod. Using $(\bfr_{i+1}-\bfr_i)=l_i\bfe_i$, we finally obtain the two forces acting on nodes:
\begin{equation}
\bff^{(t)}_{i+1}=-\bff^{(t)}_{i}=(\bfm^{(t)}_{i}/ l_i) \times \bfe_i
\end{equation}

\subsection{Contact forces distribution}\label{appendix.fc}
Let's a contact force $\bff^{(c)}$ acting at point $\bfr_{C}$. We are looking for two point forces $\bff_i^{(c)}$  (respectively $\bff_{i+1}^{(c)}$) acting at point $\bfr_i$ (resp. $\bfr_{i+1}$) and a moment $m_i\bfe_i$ such that:
\begin{subequations}
\begin{align}
\bff_i^{(c)}+\bff_{i+1}^{(c)}&=\bff^{(c)}\label{eq.split.fca}\\
(\bfr_{i+1}-\bfr_i)\times\bff_{i+1}^{(c)}+m_i^{(c)}\bfe_i&=(\bfr_{C}-\bfr_i)\times\bff^{(c)}\label{eq.split.fcb}
\end{align}
\end{subequations}

Scalar product of \eqref{eq.split.fcb} with $\bfe_i$ gives:
\begin{equation}
m_i^{(c)}=\bigl[(\bfr_C-\bfr_i)\times\bff^{(c)}\bigr]\cdot\bfe_i
\end{equation}
and cross product of \eqref{eq.split.fcb} with $\bfe_i$ gives:

\begin{align}
l_i \bff_{i+1}^{(c)}-l_i \bigl[\bff_{i+1}^{(c)}\cdot\bfe_i\bigl]\bfe_i&=
\bff^{(c)}\bigl[\bfe_i\cdot(\bfr_C-\bfr_i)\bigr]\nonumber\\
&-(\bfr_C-\bfr_i)\bigl[\bfe_i\cdot \bff^{(c)}\bigl]
\end{align}

Defining the parallel and perpendicular component of a force $\bff$ with respect to the cylinder axis as:
\begin{subequations}
\begin{align}
\bff^\parallel&=\bigr[\bff\cdot\bfe_i\bigr]\bfe_i\\
\bff^\perp&=\bff-\bff^\parallel
\end{align}
\end{subequations}
we obtain:
\begin{equation}
\bff_{i+1}^{(c),\perp}=s_i \bff^{(c),\perp}-\frac{R}{l_i}f^{(c),\parallel}\bfn\label{eq.fic}
\end{equation}
\eqref{eq.fic} determines only the components of $\bff_{i+1}^{(c)}$ which are perpendicular to the axis. The parallel component of $\bff_{i+1}^{(c)}$ is obtained in the following way. Consider the cylinder of length $l_i$ made of an elastic material, and let's $k$ the stiffness of the corresponding compressing spring. This cylinder may be viewed as the reunion of one cylinder of length $s_i l_i$ with stiffness $k/s_i$, and one cylinder of length $(1-s_i) l_i$ with stiffness $k/(1-s_i)$. Let's a force $\bff^{(c),\parallel}$ applied at the junction between cylinders. This force moves the junction on a distance $\delta=\| \bff^{(c),\parallel} \| / [k/s_i+k/(1-s_i)]$. This displacement deforms the part of length $(1-s_i) l_i$ and generates a force
$f_{i+1}^{(c),\parallel}=[k/s_i]\delta=s_i f^{(c),\parallel}$ on this spring. Inserting this equation in \eqref{eq.fic}, we finally obtain:
\begin{equation}
\bff_{i+1}^{(c)}=s_i \bff^{(c)}-\frac{R}{l_i}\bigl[f^{(c)}\cdot \bfe_i\bigr]\bfn
\end{equation}

\subsection{Distance}\label{appendix.distance}

\begin{figure}[t]
\includegraphics[width=\columnwidth]{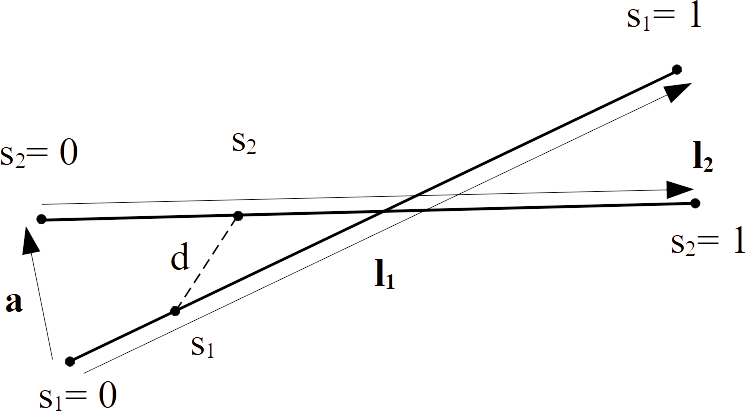}
\caption{Distance between two points located on two segments of line.}
\label{fig.distance}
\end{figure}

We consider two segments $1$ and $2$ whose axis are drawn on figure~\ref{fig.distance}. On each axis are located at abscissa $s=0$ a sphere of rayon $r$, and a segment of cylinder of radius $r$ for $0\le s \le 1$. The distance between two points at abscissa $s_1$ and $s_2$ is:

\begin{equation}
d^2(s_1,s_2)=(\bfa+s_1~\bfl_1+s_2~\bfl_2)^2
\end{equation}

The distance is minimal for $s_1^*$ and $s_2^*$ which verify:

\begin{equation}
\Bigl(\frac {\partial d^2(s_1,s_2)}{\partial s_1}\Bigr)(s_1^*,s_2^*)=\Bigl(\frac {\partial d^2(s_1,s_2)}{\partial s_1}\Bigr)(s_1^*,s_2^*)=0~\label{dmin}
\end{equation}

Equation~\ref{dmin} is solved to obtain $(s_1^*,s_2^*)$, and the minimal distance $d(s_1^*,s_2^*)$ is obtained. If $d(s_1^*,s_2^*)<2r$, with $0\le s_1^* \le 1$ and  $0\le s_2^* \le 1$, the contact is found between the two cylinders.

It not, the contact is checked between the sphere located at $s_1=0$ and the cylinder $2$. For this the minimal distance is obtained for $s_2^*$ verifying:

\begin{equation}
\Bigl(\frac {\partial d^2(0,s_2)}{\partial s_2}\Bigr)(0,s_2^*)=0~\label{dmin2}
\end{equation}

Equation~\ref{dmin2} is solved to obtain $s_2^*$, and the minimal distance $d(0,s_2^*)$ is obtained. If $d(0 ,s_2^*)<2r$, with  $0\le s_2^* \le 1$, the contact is found between the sphere (1) and the cylinder (2).

The contact between sphere (2) and cylinder (1) is searched in a similar way. If not, we check for a contact between the two spheres.


\end{document}